# GALILEO BETWEEN JESUITS: THE FAULT IS IN THE STARS

Christopher M. Graney*

ABSTRACT: In the middle of the seventeenth century, André Tacquet, S.J. briefly discussed a scientific argument regarding the structure of a Copernican universe, and commented on Galileo Galilei's discussion of that same argument—Galileo's discussion in turn being a commentary on a version of the argument by Christoph Scheiner, S.J. The argument was based on observations of the sizes of stars. This exchange involving Galileo and two Jesuits illustrates how through much of the seventeenth century, science—meaning observations measurements, and calculations—supported a view of the Copernican universe in which stars were not other suns, but were dim bodies, far larger than the sun. Johannes Kepler emphasized this, especially in arguing against Giordano Bruno. Jesuit astronomers like Tacquet and Scheiner understood this. Those who might have listened to Jesuit astronomers would likewise have understood this—Robert Bellarmine, for example, whose role in the debate over Copernicanism is well known. To many, such a universe was, in the words of Galileo's Dialogue character Sagredo, "beyond belief," and no modern view of a universe of many distant suns would be scientifically supportable until after Tacquet's death in 1660. The Copernican universe of the seventeenth century looked radically different from the universe as modern astronomers understand it, and recognizing this fact allows for interesting questions to be asked regarding the actions of those, such as Bellarmine, who were responding to the work of Copernicus.

What did the Copernican universe look like to Jesuit astronomers in the seventeenth century?  In the middle of that century, André Tacquet of the Society of Jesus produced an unusual and compact version of a key scientific argument regarding the structure of a Copernican universe.  He also briefly commented on Galileo Galilei's lengthy discussion of that same key argument.  Galileo's discussion was itself a commentary on a different, likewise compact version of the argument, produced earlier in the century by another Jesuit astronomer, Christoph Scheiner.  This exchange between Galileo and two Jesuit astronomers illustrates a thing not widely understood regarding the Church's interaction with Copernican ideas: that, to a Jesuit astronomer—and indeed to any astronomer persuaded by observations, measurements, and calculations—the Copernican universe looked radically different from the universe as we understand it today.  And it looked likewise, we may presume, to anyone who might have consulted Jesuit astronomers for their expertise.  Because the ideas of these astronomers had influence on the broader Church, and because the Church's interaction with Copernican ideas is a prominent aspect of its history, this exchange between Scheiner, Galileo, and Tacquet regarding the structure of a Copernican universe is discussed here in detail.

THE SIZES OF STARS

André Tacquet, S.J. lived from 1612 to 1660.  References to Tacquet are not common in recent scholarship.  Perhaps this is in part because, as G. H. W. Vanpaemel wrote in a piece on "Jesuit Science in the Spanish Netherlands," Tacquet's life "was utterly uneventful; he apparently never ventured outside the borders of his native province."  It may also be because, according to Vanpaemel, while Tacquet produced original work, much of his effort was spent on teaching mathematics, and on producing work for teaching.[1]  While Vanpaemel's assessment of Tacquet may be correct, we shall see that Tacquet became known for an argument he made against the heliocentric universe of Nicolas Copernicus.

---

*Prof. Graney is an astronomer and historian of science on the staff of the Specola Vaticana (the Vatican's astronomical observatory) in Rome and Tucson.  His email address is c.graney@vaticanobservatory.org.[1] Geert H. W. Vanpaemel, "Jesuit Science in the Spanish Netherlands," in: *Jesuit Science and the Republic of Letters*, ed. Mordechai Feingold (Cambridge, MA, 2003), 406.

Page 2 of 39

This argument was based on the apparent sizes of the stars, specifically of the "fixed" stars. Fixed stars are those we call stars today, that make up the constellations. This is as opposed to the "wandering" stars that move through the constellations over time, otherwise known as "planets" ("planet" meaning "wanderer"). Astronomers from Ptolemy in ancient times to Tycho Brahe in the late sixteenth century had attempted to measure the sizes of these stars as seen from Earth. They determined the "apparent diameters" of the more prominent or "first magnitude" (i.e. "first rate") fixed stars to be roughly one fifteenth the apparent diameter of the moon; fifteen such stars, placed one against another in a straight line, would equal the apparent diameter of the full moon. The wandering stars had similar apparent diameters.[2]

In a heliocentric universe, Earth's position relative to any given fixed star necessarily changes as Earth moves annually around the sun. As Earth circles through its orbit, the distance from Earth to that star changes, and the angle of view toward that star also changes, owing to Earth's continually changing location with respect to the star. These changes must produce observable annual variations in the appearance of the star, variations known as "annual parallax." No such variations were observed in any fixed star, in Copernicus's time or in Tacquet's. This meant that Earth's orbit had to be vanishingly small compared to the distance of the fixed stars. But for the fixed stars to be at such large distances yet still show measurable apparent diameters required that their actual diameters be enormous (Figure 1). Every visible star in a heliocentric universe had to be a body far larger than the sun.

By contrast, in a geocentric universe the Earth was immobile, so there was no expectation of annual parallax. Stars could therefore be located just beyond Saturn. Being comparable to Saturn in apparent size, and lying at a distance comparable to that of Saturn, stars would have actual sizes also comparable to Saturn. In a geocentric universe, fixed stars were commensurate in size with the other celestial bodies.

Thus the Copernican hypothesis turned the fixed stars into a new class of giant celestial bodies, far different from the earth, sun, moon, and planets. The question of Copernican star sizes had been raised by Brahe, who found the idea of enormous stars too improbable, and considered them an argument against heliocentrism.[3] This challenge to Copernicus's ideas is not widely understood today, but as Albert Van Helden has written, "Tycho's logic was impeccable; his measurements above reproach. A Copernican simply had to accept the results of this argument" and agree that the stars were giant.[4] Vanpaemel writes that Tacquet elaborated on this argument:

> Tacquet proved that in the Copernican hypothesis the proportion of the dimensions of the fixed stars to the distance earth-sun, would be equal to the proportion of the dimensions of the same stars to the radius of the earth in the geocentric hypothesis. In the Copernican hypothesis therefore, the stars needed to be much larger and heavier than in the traditional view, a conclusion which conflicted with intellectual economy.[5]

---

[2] Albert Van Helden, *Measuring the Universe: Cosmic Dimensions from Aristarchus to Halley* (Chicago, 1985), 27, 30, 32, 50.

Brahe had proposed his own hypothesis, in which the sun, moon and stars circled an immobile Earth while the planets circled the sun (Figure 2). This system was identical to the Copernican system insofar as any Earth-bound observation of the sun, moon, or planets was concerned; later telescopic discoveries involving those bodies, such as the phases of Venus that showed it to orbit the sun, were thus fully compatible with Brahe's system.

THE *DISQUISITIONS* OF SCHEINER AND LOCHER

Brahe may have first raised the star-size question against heliocentrism, but the Jesuit astronomer Christopher Scheiner and his student Johann Georg Locher arguably wrote the most consequential discussion of it. In their 1614 book *Mathematical Disquistions*, Scheiner and Locher spent a few pages outlining the issue by means of calculations.[6] Then they produced this very compact and clear version of Brahe's idea:

> [In a Copernican universe] even the smallest star visible to the eye is much larger than the whole circle of Earth's orbit. This is because such a star has a measurable size, as does the circumference of the sky. The ratio of the size of the star to the size of the firmament of fixed stars is therefore perceptible. But according to the Copernican opinion, the ratio of the size of the circle of Earth's orbit to the size of the firmament is imperceptible. For in the Copernican opinion the size of the Earth's orbital circle holds the same proportion to the firmament as the size of Earth itself holds to the firmament in the common geocentric opinion. Yet experience shows the Earth to be of imperceptible size compared to the firmament. Thus in the Copernican opinion it is the circle of Earth's orbit that is of imperceptible size compared to the firmament—and therefore smaller than the smallest perceptible star.[7]

---

[3] Christopher M. Graney, *Setting Aside All Authority: Giovanni Battista Riccioli and the Science Against Copernicus in the Age of Galileo* (Notre Dame, Indiana, 2015), 25-44.
[4] Van Helden, *Measuring the Universe*, 51.
[5] Vanpaemel, "Jesuit Science," 409.
[6] Christopher M. Graney, *Mathematical Disquisitions: The Booklet of Theses Immortalized by Galileo* (Notre Dame, Indiana, 2015), 27-31.

Scheiner and Locher thus offered a simple scientific (that is, based on observation, measurement, and calculation) argument that in a Copernican universe every last visible star must necessarily have an actual physical size not just exceeding the sun, but exceeding Earth's orbit around the sun. This meant every last visible star would dramatically dwarf the sun, and every other celestial body, more than a beach ball dwarfs a pea. Their logic can be distilled down to this: if Earth's orbit is vanishingly small in a Copernican universe, while the fixed stars have small but measurable apparent sizes, then it follows that every visible star must be larger than Earth's orbit—because "small but measurable" is larger than "vanishingly small."

This brief discussion was consequential because two decades later Galileo Galilei would devote significant space in his 1632 *Dialogue Concerning the Two Chief Worlds Systems—Copernican and Ptolemaic* to addressing the contents of *Disquisitions*, with many pages spent in response to its discussion of star sizes. Thus Galileo immortalized *Disquisitions* within the *Dialogue*, one of the most famous books in the history of science and of the Church.

None of the characters in the *Dialogue* embraced the idea of enormous stars, but some Copernicans actually did. Notable among these was Johannes Kepler. He saw no issues of intellectual economy in these stars. He agreed that all visible stars were larger than Earth's orbit. Indeed, he estimated the most prominent ones to be larger than Saturn's orbit, and thus larger than an entire geocentric universe. Kepler also reasoned that the fixed stars were dim. This was because their combined power to illuminate the sky was insignificant compared to the sun, while the fraction of the sky that they together occupied was comparable to that occupied by the sun. Their distance did not excuse their dimness, Kepler said, because the farther away they were, the larger they would be (as seen in Figure 1).

---

[7] Graney, *Disquisitions*, 30.

Thus observations, measurements, and calculations proved that the universe consisted of myriad vast, dim distant stars enveloping a single, unique sun (itself both tiny and brilliant compared to them) and the sun's still tinier planets. Kepler attacked Giordano Bruno's notion that the stars were other suns, orbited by other Earths; what Bruno said was contrary to what any astronomer could easily determine for himself.[8] Kepler argued in his 1606 *De Stella Nova* that the brilliant, tiny sun versus the dim, hulking stars spoke to God's ability to create on a vast scale while still retaining full concern for the smallest things.[9] Scheiner and Locher, perhaps with Kepler in mind, remarked in *Disquisitions* that Copernicus's "minions" did not deny the enormous stars, and instead "go on about how from this everyone may better perceive the majesty of the Creator," an idea they judged to be "laughable."[10]

TACQUET ON STARS AND GALILEO

Tacquet's discussion of the star size question includes no such judgements. It is found in his posthumously published *Opera Omnia*, Book 5 ("Concerning the Fixed Stars"), Chapter 2 ("Concerning the magnitude, distance, light, number, kinds, and forms of the Fixed Stars").[11] Book 5 is divided into numbered sections. The first section in Chapter 2 is number 21. Here Tacquet discusses what Earth-bound astronomers can determine regarding distances to the stars in a geocentric universe. In a geocentric universe, astronomers are limited to observing celestial bodies from different places on Earth's surface, one Earth radius (or semidiameter) from the center of the geocentric universe. By observing a celestial body from those different places on Earth's surface (Figure 3), they can measure angles to that body from those places, and triangulate to determine the distance of that body. This is using an Earth-surface "parallax" to find distance.

---

[8] Christopher M. Graney, "The Starry Universe of Johannes Kepler," *Journal for the History of Astronomy,* 50 (2019), 162-66.
[9] Christopher M. Graney, "As Big as a Universe: Johannes Kepler on the Immensities of Stars and of Divine Power," *Catholic Historical Review,* 105 (2019), 75-90.
[10] Graney, *Disquisitions*, 29
[11] André Tacquet, *Opera Mathematica* (Antverpiæ: Apud Iacobum Meursium, 1668), 205.

Triangulating from observations of the moon made at different places on Earth's surface provided for accurate determination of its distance, for example. However, the more distant the object, the narrower the key angle at the celestial object (at 'D' or 'C' in Figure 3). When the object is too distant, that angle becomes too small to measure; then there can be no surface parallax detected, and no distance determined.

Tacquet notes that no surface parallax has ever been detected for the fixed stars. But, he says, that does not mean that we know nothing of their distance. Rather, we know the lower limit of their distance—they must be farther away than that minimum distance within which we could just detect a parallax. He writes,

> Therefore a limit of distance is recognized (which itself is indeed an extraordinary monument to Astronomy); it is certain the Fixeds may not lie within it. The other limit of distance, beyond which they may not lie, is truly not able to be explored by human ingenuity. It may be the Fixed Stars are as much distant from Saturn, as Saturn is from us. It is even credible that some are higher than others; which perhaps is the reason why they may appear unequal to us.[12]

This last idea about how, owing to differing distances, stars appear "unequal"—not all the same in apparent size—had been discussed by St. Albert the Great:

> *We ourselves have indeed recognized one star to be more or less distant than another through the greater or lesser [apparent] diameter of the stars: but we are not able to distinguish the number of stars....*[13]

In number 22 Tacquet discusses the speeds of the stars as they circle Earth in a geocentric universe. In number 23 he discusses their apparent sizes. He reproduces a table, from the 1651 *New Almagest* of the Italian Jesuit Giovanni Battista Riccioli, listing the apparent diameters of twenty-one stars, measured via telescope.[14] In this table mighty Sirius, the "Dog Star," the most prominent fixed star in the sky, measures 18 "seconds of arc." That is 1/100$^{th}$ the apparent diameter of the moon, which is half of one degree, or 1800 seconds of arc. In the same table, insignificant Alcor (a star in the handle of the Big Dipper, known mostly for being close to the star Mizar that marks the bend of that handle) has an apparent diameter of just over 4 seconds of arc. These sizes are much smaller than what Ptolemy or Brahe would have recognized. Recall that they had both determined that prominent stars measured about a fifteenth of the moon's apparent diameter, roughly 120 seconds of arc. The advent of the telescope had prompted a reassessment of the apparent diameters of both fixed and wandering stars.

---

[12] Tacquet, *Opera Mathematica,* 205: "Notus est igitur (quod ipsum quidem monumentum Astronomiae eximium est) distantiae terminus, infra quem Fixas non deprimi certum sit. Terminus vero distantiae alter supra quem non ascendant, explorari humano ingenio non potest. Fortassis Stellae Fixae tantum a Saturno distant, quantum ille a nobis. Credibile est etiam alias esse alijs altiores; quae & forsan est causa, cur inaequales nobis appareant."

[13] Albertus Magnus, *Commentarii in II Sententiarum*, (Luguduni: Prost, 1651). Dist. 2, Art. 3, "Ad Quaest.": "nos quidem deprehenderemus unam stellam magis vel minus distare quam aliam per diametrum stellarum magis vel minus: sed non possumus distinguere numerum stellarum...."

In number 24 Tacquet uses Riccioli's table of apparent stellar diameters, and an estimated minimum distance allowable for the stars to show no Earth-surface parallax, to produce a table of actual sizes for thirteen stars, again for a geocentric universe. Tacquet's result is that Sirius is a minimum of 815 times greater than Earth as measured by volume; Alcor is nine times greater.[15] This would make Sirius a body whose actual diameter was at minimum nine times that of the Earth, and Alcor a body at minimum twice Earth's diameter. For comparison, elsewhere Tacquet determines the sun's actual diameter to be roughly twelves times that of Earth.[16] Thus Sirius and Alcor are larger than the Earth, and smaller than the sun.

Number 25 treats the volume of the geocentric universe. This discussion assumes the stars are all equidistant from Earth, even though number 21 discussed the possibility that they could lie at various distances past Saturn.

At number 26 Tacquet arrives at the Copernican hypothesis. He points out that the Earth's orbit in a heliocentric universe holds the same position as the Earth itself in a geocentric universe:

---

[14] See Tacquet, *Opera Mathematica,* 206, and Giovanni Battista Riccioli, *Almagestum Novum* (Bononinae : Ex typographia Haeredis Victorii Benatii, 1651), vol. 1, 716.

[15] There are various errors in Tacquet's calculations, but these do not change the overall picture significantly—see Christopher M. Graney, "Not the Earth, but its Orbit: André Tacquet and the Question of Star Sizes in a heliocentric universe," *arXiv:1909.12074v1* (Sept. 24, 2019), 1-16, here 10-11.

[16] Tacquet, *Opera Mathematica,* 143.

*In the hypothesis of a Moved Earth, the Great Orb (that is, the sphere whose semidiameter is the distance from the Sun to our eye or to the center of the Earth) is, compared to the Firmament [of the Fixed Stars], the equivalent of a point.* No [Earth-surface] Parallax in the Fixed Stars has ever been detected, even though such Parallax ought to arise, owing to the distance of our eye from the center of the Firmament [being one semidiameter of Earth]. Therefore it is evident that distance [of one terrestrial semidiameter] is insensible compared to the distance of the Fixeds. Yet in the hypothesis of the Moved Earth (that is, where the Sun may rest in the center of the Firmament, with the Earth running through the Ecliptic), the distance of our eye from the center of the Firmament is the distance of the eye from the Sun. This is the semidiameter of the Great Orb. Therefore in the Hypothesis of the Moved Earth, the semidiameter of the Great Orb is insensible compared to the distance of the Fixeds, or the semidiameter of the Firmament. Whereby the proportion of spheres may be the cube of the proportion of diameters, *a fortiori* (as the Philosophers say), the Great Orb itself will be insensible compared to the Firmament.

*Corollary: Hence it follows that the Great Orb is to the Copernican Firmament (that required by the Hypothesis of the Moved Earth) as the Earth is to the common Firmament (that required by the Hypothesis of the Standing Earth).*[17]

In other words, in a heliocentric universe, as Earth travels around the sun on its annual orbit, astronomers can observe celestial bodies from different places along that orbit, one orbital radius (or semidiameter) from the sun, the center of the heliocentric universe. As no parallax of any sort is seen in the fixed stars, the semidiameter of Earth's orbit in a heliocentric universe is just like the semidiameter of Earth in a geocentric universe—vanishingly small compared to the distance to the fixed stars.

---

[17] Tacquet, *Opera Mathematica,* 208: "*In Hypothesi Terrae Motae, Orbis Magnus, sive Sphaera cuius semidiameter est distantia oculi nostri vel centri Terrae a Sole, ad Firmamentum instar puncti est.* Cum enim in Stellis Fixis numquam vlla Parallaxis fuerit deprehensa, Parallaxis autem illa siqua esset, oriri deberet ab oculi nostri distantia a centro Firmamenti; perspicuum est distantiam illam ad distantiam Fixarum esse insensibilem. Atqui in Hypothesi Terrae Motae, (hoc est, si Terra Eclipticam percurrente, Sol in Firmamenti centro quiescat) distantia oculi nostri a centro Firmamenti, est ipsa distantia oculi a Sole, hoc est Orbis Magni semidiameter. Ergo in Hypothesi Terrae Motae semidiameter Orbis Magni ad distantiam Fixarum, siue Firmamenti semidiametrum, insensibilis est. Quare cum sphaerarum proportio triplicata sit proportionis diametrorum, ipse Orbis Magnus ad Firmamentum a fortiori (vt loquuntur Philosophi) insensibilis erit. *Corollarium: Hinc sequitur Orbem Magnum esse ad Firmamentum Copernicanum, hoc est debitum Hypothesi Terrae Motae; vt Terra est ad Firmamentum commune; hoc est, debitum Hypothesi Terrae Stantis.*"

After running through a more rigorous geometrical demonstration of these statements, Tacquet proceeds to number 27—his own version of the star size argument:

> *In the Hypothesis of the Moved Earth, the Fixed Stars hold the same proportion to the Great Orb, that they hold to the Earth in the Hypothesis of the Standing Earth.* The Apparent Diameter of any Fixed (that is, the angle under which the Fixed is seen from Earth), may be found independently from either Hypothesis of the Earth (Moved or Standing). As is obvious from *number 23*, it is clear a Fixed (*Spica*, for example) to subtend in either Hypothesis equally many Seconds of its Firmament; obviously just that many, as are recorded by observation. Therefore the Copernican *Spica* is to the Copernican Firmament, as the common [geocentric] *Spica* is to the common [geocentric] Firmament. And yet the Copernican Firmament is to the Great Orb, as the common Firmament is to the Earth. Thus the Copernican *Spica* is to the Great Orb, as the common *Spica* is to the Earth, by reason of proportion.[18]

In number 28, Tacquet points out that, since in the geocentric hypothesis (recall that Brahe's geocentric hypothesis was compatible with telescopic discoveries[19]) Sirius is 815 times and Alcor 9 times greater than Earth by volume, it follows that in the Copernican hypothesis Sirius and Alcor will be 815 and 9 times greater by volume than the Great Orb—that is, than the sphere of Earth's orbit. Thus, in terms of actual diameter Sirius and Alcor will be nine times and twice, respectively, Earth's orbit—at minimum. Tacquet closes number 28 with the following:

---

[18] Tacquet, *Opera Mathematica,* 208: "*In Hypothesi Terrae Motae, Stellae Fixae eam habent proportionem ad Orbem Magnum, quam habent in Terrae Stantis Hypothesi ad Terram.* Cum Diameter Apparens cuiusuis Fixae, hoc est angulus sub quo Fixa videtur ex Terra, reperiatur praescindendo ab vtralibet Hypothesi Telluris Motae vel Stantis, vt patet ex *nu.* 23, liquet Fixam, ex. gr. *Spicam*, in vtraque Hypothesi aeque multa sui Firmamenti Minutae subtendere; tot nimirum, quot obseruatio exhibet. Igitur *Spica* Copernicaea est ad Firmamentum Copernicaeum, vt *Spica* communis est ad Firmamentum commune. Atqui Firmamentum Copernicaeum est ad Orbem Magnum, vt Firmamentum commune ad Terram. Ex aequo igitur *Spica* Copernicaea est ad Orbem Magnum, vt *Spica* communis ad Terram." Note that here I have taken Tacquet's "Firmamenti Minutae" as having been a typographical error that should have read "seconds", since in Number 23 the apparent stellar diameters are given in seconds, not minutes.

[19] See Tacquet, *Opera Mathematica,* 258 for examples of his using Brahe's hypothesis with Venus.

Galileo in his System of the World attempts in vain to elude this monstrous magnitude of the Fixeds by a long discourse, from page 350 up to 383.[20] But he weakens nothing of what we have demonstrated through *numbers 26-28*.[21]

This ends Tacquet's star size discussion. Like Scheiner and Locher, he used just a few pages.

STARS UNCHANGED

Often we think of heliocentrism in terms of making the starry universe larger—pushing the stars farther away, much like Rheticus accuses Copernicus of doing in Dava Sobel's play *And the Sun Stood Still*: "The stars get in your way? You just wave them off to some other place."[22] But under a seventeenth-century understanding of stars, pushing them farther away enlarges them proportionately, via Figure 1. Their separations relative to their sizes remain unaltered. This fact can elude us, with the result that we imagine that pushing the stars out yields a heliocentric universe of stars that is vaster and emptier than its geocentric counterpart. But on a stellar scale, the heliocentric universe of stars is no vaster nor emptier than the geocentric universe; both are identical, under a seventeenth-century understanding.

Tacquet's unusual discussion invites us to think of heliocentrism not as making the starry universe *larger*, but as making the earth, sun, moon, planets, and their orbits within it all *smaller*, as the heliocentric Earth's orbit replaces the geocentric Earth. If Earth is made smaller, then on any Earth-based scale the universe will *seem* enlarged, but in fact the starry universe remains unchanged. As Tacquet notes, the stars occupy the same proportion of space within the sky, whether the universe is heliocentric or geocentric. But, in the switch from geocentrism to heliocentrism, the Earth and sun are dramatically reduced; stars that were smaller than the sun in a geocentric universe dwarf it in a heliocentric one. Thus not only does mighty Sirius utterly dwarf the heliocentric universe's sun, but little Alcor does, too.

---

[20] Galileo Galilei, *Dialogue Concerning the Two Chief World Systems: Ptolemaic and Copernican*, trans. Stillman Drake [Modern Library Science] (New York, 2001), 416-52 – hereafter cited as *Dialogue*.

[21] Tacquet, *Opera Mathematica,* 209: "Galilaeus in suo Mundi Systemate immanem istam Fixarum magnitudinem nequidquam conatur eludere a pag. 350 *vsque ad* 383 discursu longissimo, sed nihil eorum, quae *num.* 26, 27, 28, iam demonstrauimus, infirmante."

[22] Dava Sobel, *A More Perfect Heaven: How Copernicus Revolutionised the Cosmos* (London, 2011), 128.

Like Scheiner and Locher's summary of the star size issue, Tacquet's is compact, clear, and scientific (that is, based on observations, measurements, and calculations). The specific details of sizes are not important—what Tacquet says holds so long as no parallax is detected, and stars show measurable size. Now let us consider the star size discourse that he so briefly dismissed—the lengthy one found in Galileo's *Dialogue*.

SIMPLICIO, SAGREDO, SALVIATI, AND STARS

Galileo discusses the star size question on the "Third Day" of his *Dialogue*. The character Sagredo opens the door for the discussion, saying that "it is now time for us to hear the other [anti-Copernican] side, from that booklet of theses or disquisitions which Simplicio has brought back with him."[23] That booklet is Scheiner and Locher's *Disquisitions*. It had already been invoked by the character Simplicio on the first and second days,[24] and roundly ridiculed by the character Salviati. Here on the third day, Salviati again ridicules it, even making reference to certain "apish puerilities" contained within.[25] Then Simplicio brings the discussion around to what is in *Disquisitions* regarding the star size question:

> Now here, as you see, he [Scheiner/Locher] deduces… that if the orbit in which Copernicus makes the earth travel around the sun in a year were scarcely perceptible with respect to the immensity of the stellar sphere, as Copernicus says must be assumed, then one would have to declare and maintain that the fixed stars were at an inconceivable distance from us, and that the smallest of them would be much larger than this whole orbit…[26]

Salviati replies that this argument is based on the introduction of "false assumptions," and states that,

> by assuming that a star of the sixth magnitude may be no larger than the sun, one may deduce by means of correct demonstrations that the distance of the fixed

---

[23] *Dialogue*, 414.
[24] *Dialogue*, 105, 253.
[25] *Dialogue*, 415.
[26] *Dialogue*, 416.

stars from us is sufficiently great to make quite imperceptible in them the annual movement of the earth…[27]

Salviati then notes that the apparent diameter of the sun measures one half of one degree—1800 seconds of arc, like the moon. First magnitude fixed stars have an apparent diameter measuring no more than five seconds of arc, he says, while fixed stars of the sixth magnitude (those barely visible to the eye) measure 5/6 of a second of arc.[28] Thus the apparent diameter of the sun is $1800 \div (5/6) = 2160$ times greater than that of a sixth magnitude star. And so, Salviati continues,

> if one assumes that a fixed star of the sixth magnitude is really equal to the sun and not larger, this amounts to saying that if the sun moved away until its diameter looked to be $1/2160^{th}$ of what it now appears to be, its distance would have to be 2160 times what it is now.[29]

So, were a star of the sixth magnitude equal in size to the sun, it would lie at a distance of 2160 solar distances, and the effect of earth's orbit "would be little more noticeable than that which is observed in the sun due to the radius of the earth."[30]

Salviati goes on at some length about erroneous assumptions regarding the apparent diameters of stars—the false assumptions he mentioned above. Earlier astronomers who determined that a prominent star appears to have an apparent diameter about one fifteenth that of the moon (120 seconds of arc) cannot be excused, he says, for their erroneous measurements of the stars (both fixed and wandering). This is because

> it was within their power to see the bare stars at their pleasure, for it suffices to look at them when they first appear in the evening, or just before they vanish at dawn. And Venus, if nothing else, should have warned them of their mistake, being frequently seen in daytime so small that it takes sharp eyesight to see it, though in the following night it appears like a great torch.[31]

---

[27] *Dialogue*, 417.
[28] When Galileo first reported in 1610 on the appearance of fixed stars as seen through a telescope, he described them as not having an obviously round appearance. However, by 1613 he was describing them as clearly round, and did so repeatedly from then onward, always ascribing to them an apparent diameter of a few seconds of arc. See Graney, *Setting Aside*, 46-48.
[29] *Dialogue*, 417.
[30] *Dialogue*, 418.
[31] *Dialogue*, 419.

Salviati notes that the telescope, "by showing the disc of the star bare and very many times enlarged," renders the process of measuring apparent diameters of fixed and wandering stars much easier (Figure 4).[32] However, he argues that non-telescopic methods can also remove the "adventitious irradiation" from a fixed or wandering star (such as Venus) and reveal its *correct* apparent diameter; these methods again show that to be no more than five seconds in the case of a first magnitude star.[33] The five second value is smaller than the apparent diameter Riccioli would later determine for Sirius, although it overlaps with Riccioli's measurement for stars like Alcor. Salviatti goes on to spend several pages discussing the non-telescopic methods.

However, 2160 solar distances is 216 times the distance of Saturn. Brahe, having searched for parallax but not detected it, stated that fixed stars would have to be at least 700 times more distant than Saturn.[34] Thus Simplicio comes back to the failure to detect annual parallax; 2160 solar distances is not far enough to explain its absence:

> when one assumes the star of the sixth magnitude to be as large as the sun… it still remains true that the earth's orbit would necessarily cause changes and variations in the stellar sphere similar to the observable changes produced by the earth's radius in regard to the sun. No such changes, or even smaller ones, being observed among the fixed stars, it appears to me that by this fact the annual movement of the earth is rendered untenable and is overthrown.[35]

Salviati responds that "nothing prevents our supposing that the distance of the fixed stars is still much greater than has been assumed." He argues that the periods of celestial bodies suggest that the stars could be much more distant still. He notes how, in the Copernican system, the periods of time required for celestial bodies to complete their cycles increase with distance from the sun: Saturn is farther from the sun than Jupiter and takes longer than Jupiter to circle the sun; likewise for Jupiter versus Mars. Salviati points out that, according to Ptolemy, the period for the precession of the fixed stars is over a thousand times longer than Saturn's period. By comparing the period of the fixed stars to those of Saturn, Jupiter, and Mars, Salviati supposes that stellar distances could be five, seven, or twelve times larger, respectively, than the 2160 solar distance value he previously calculated.[36]

---

[32] *Dialogue*, 419.
[33] *Dialogue*, 421.
[34] Ann Blair, "Tycho Brahe's Critique of Copernicus and the Copernican System," *Journal for the History of Ideas,* 51 (1990), 355-77, here 364; Kristian Peder Moesgaard, "Copernican Influence on Tycho Brahe," in: *The Reception of Copernicus' Heliocentric Theory*, ed. Jerzy Dobrzyck (Dordrecht, 1972), 30-55, here 51.
[35] *Dialogue*, 423.

But to suppose the stars are five, seven, or twelve times more distant than previously calculated is to suppose (via Figure 1) that the stars are five, seven, or twelve times greater in actual diameter than previously calculated—namely, that sixth-magnitude stars are five, seven, or twelve times greater in actual diameter than the sun (and thus 125, 343, or 1728 times greater in volume). The more prominent stars, such as first-magnitude stars, are larger still; Salviati had said these measure six times the apparent diameter of the sixth-magnitude stars. If first- and sixth-magnitude stars are assumed to be similarly distant, then the first-magnitude stars must be six times the actual diameter of the sixth-magnitude stars. First- and sixth-magnitude stars cannot all be assumed to be of the same size, for that would require the first-magnitude ones to be six times closer, and thus subject to greater parallax.

Salviati and Sagredo at this point launch into a discussion regarding Divine Providence, what sizes are comprehensible, whether some size is greater than God can accomplish, and whether certain amounts of space are purposeless. They discuss the relative sizes of elephants and ants, whales and gudgeons, the universe and the moon. Sagredo declares that "a great ineptitude exists on the part of those who would have it that God made the universe more in proportion to the small capacity of their reason than to His immense, His infinite, power."[37] However, Simplicio, citing *Disquisitions* again, pulls the discourse back to the star size question. Salviati has implicitly granted that little sixth-magnitude stars, barely visible to the eye, are at least five times the sun's actual diameter and 125 times the sun's volume, with prominent stars being thirty times the sun's actual diameter, and 27,000 times its volume. The question of the "other side" remains unanswered then. Notes Simplicio:

> All this that you are saying is good, but what the other side objects to is having to grant that a fixed star must be not only equal to, but much greater than, the sun;

---

[36] *Dialogue*, 424.
[37] *Dialogue*, 429.

for both are still individual bodies located within the stellar orb [i.e. within the universe].[38]

Here Simplicio is pointing out that in a geocentric universe individual fixed stars are commensurate in size to the other *individual bodies* found within the universe, while in a heliocentric universe they are not. Even under Salviati's own numbers, the Copernican system still requires the stars to become a new class of giant celestial bodies—an "ad hoc" creation to answer the scientific problem of parallax. In 1633 Galileo stated, in depositions to the Inquisition, that he did not intend for the *Dialogue* to promote the Copernican system:

> I did not do so [write the *Dialogue*] because I held Copernicus's opinion to be true. Instead, deeming only to be doing a beneficial service, I explained the physical and astronomical reasons that can be advanced for one side and for the other; I tried to show that none of these, neither those in favor of this opinion or that, had the strength of a conclusive proof....[39]

He stated that he did not hold the Copernican opinion at that time, did not hold it when he wrote the *Dialogue*, and that he simply over-wrote the pro-Copernican side of the book (to an extent he had not realized until he had recently reread the book) merely out of a desire to appear clever.[40] Simplicio's incisive "individual bodies" comment regarding the star size question might have been something in the *Dialogue* to which Galileo could have referred in support of these statements to the Inquisition, had he wished to do so.

Within the *Dialogue*, however, Simplicio's comment is allowed to pass. The conversation quickly reverts to discussions of "purpose." Then Salviati declares that he has "already demonstrated two things." One of these is the distance to the fixed stars needed to solve the parallax problem (he ignores that Simplicio has said that he has not). The other is that a visible "fixed star" need not be assumed to be larger than the sun (he here omits reference to just "sixth-magnitude" fixed stars). He then proceeds to question whether anyone has actually "tried to investigate in any way whether any phenomena is perceived in the stellar sphere by which one might boldly affirm or deny the annual motion of the earth."⁴¹ Sagredo answers "no," that they would have no need, since Copernicus already says no parallax can be detected. Sagredo continues:

> Then on this assumption they show the improbability which follows from it; namely, it would be required to make the [stellar] sphere so immense that in order for a fixed star to look as large as it does, it would actually have to be so immense in bulk as to exceed the earth's orbit—a thing which is, as they say, entirely unbelievable.⁴²

---

[38] *Dialogue*, 430.
[39] Maurice Finocchiaro, *The Galileo Affair: A Documentary History* (Berkeley, 1989), 287.
[40] Finocchiaro, *Galileo Affair*, 278, 287.

*Here Sagredo agrees with Brahe that enormous fixed stars are in fact entirely unbelievable.* Salviati may speak of elephants and ants, whales and gudgeons, but Sagredo apparently considers having the sun be the sole ant in a universe of elephants, the sole gudgeon in a universe of whales, to be beyond belief. Despite his earlier statement about seeing a great ineptitude in those who put limits on the power of God, Sagredo seems to think intellectual economy does imply some constraints. Kepler, of course, would have disagreed.

Salviati passes over Sagredo's "entirely unbelievable" remark, focusing instead on Sagredo's answer of "no." Salviati says that arguments based on Copernicus's own statements "may suffice to refute the man, but certainly not to clear up the fact."⁴³ He now turns to methods by which annual parallax might be detected and Earth's orbital motion revealed, engaging in a lengthy discussion with Sagredo regarding how fixed stars in different regions of the sky will show this or that specific effect or variation owing to that motion.

During this discussion, Simplicio reminds the others that, regardless of the details, he feels repugnance at being asked to grant that stellar distances are so great that all the variations discussed will be "entirely imperceptible." He notes that "if the variation is null, then the annual motion attributed to the Earth in its orbit is null." He grants that, were those variations detectable, he would have to concede the point to Salviati.⁴⁴

Salviati, meanwhile, suggests two methods for detecting the variations in the stars caused by Earth's annual motion. He believes these methods have not been tried before and would be more sensitive than those that have been tried. They would be more sensitive even than what Tycho Brahe could achieve with his instruments—and Salviati grants that Brahe's vaunted instruments were expensive and his skill at observations remarkable.

---

⁴¹ *Dialogue*, 431-32.
⁴² *Dialogue*, 432.
⁴³ *Dialogue*, 432.
⁴⁴ *Dialogue*, 448, 438, 449.

Salviati's first suggestion is to use the fixed stars by themselves to reveal Earth's motion. He states,

> it is not entirely impossible for something some time to become observable among the fixed stars by which it might be discovered what the annual motion does reside in. Then they, too... would appear in court to give witness to such motion in favor of the earth. For I do not believe that the stars are spread over a spherical surface at equal distances from one center; I suppose their distances from us vary so much that some are two or three times as remote as others. Thus if some tiny star were found by the telescope quite close to some of the larger ones, and if that one were therefore very very remote, it might happen that some sensible alterations would take place among them [that reflect Earth's annual motion].[45]

Salviati is suggesting that, if stars lie at differing distances (as we saw St. Albert the Great describe earlier) and not all on a sphere, then one star could be used as a reference point against another. Were two stars arranged as Salviati states—a smaller-looking star next to a larger-looking one, the smaller-looking star being farther away than the other— it would follow that, were Earth orbiting the sun, the alignment of these two stars would change over the course of a year. The stars would exhibit a "differential parallax" (Figure 5).

However, when Salviati remarks on what might occur "*if* some tiny star were found by the telescope" near a larger star, he does not mention that Galileo had found examples of such "double stars," and observed them with the telescope, a decade and a half prior to the *Dialogue's* publication. Galileo and his friend Benedetto Castelli made the first known observation of a double star in 1617. The star was Mizar, the star at the bend of the handle of the Big Dipper. Seen with the naked eye, it appears to be a single star, but a telescope reveals it to be double.

---

[45] *Dialogue*, 444.

Castelli observed Mizar first, then informed Galileo about it.[46] Galileo observed it and measured it with precision, recording both the separation of its two components and the apparent diameters of those components. Based on this record, we can reconstruct what he saw through his telescope, as seen in Figure 6. Galileo also calculated the distance to the brighter of the two Mizar component stars, which astronomers now call *Mizar A*. As in Salviati's *Dialogue* discussion of sixth-magnitude stars, Galileo assumed Mizar A to be the same actual size as the sun, and then calculated that, since he measured the apparent diameter of Mizar A as being 1/300$^{th}$ that of the sun, its distance was 300 solar distances.[47]

Thus while Salviati states that, "*if* some tiny star were found by the telescope" near a larger one, then the stars would "appear in court to give witness" to Earth's motion, in fact such a star had been found in Mizar in 1617. Galileo is known to have observed other double or multiple star systems, including the Trapezium system in Orion, which he sketched with great accuracy, all more than a decade prior to publication of the *Dialogue*.[48] Mizar shows no differential parallax, even though the differential parallax for a double star at 300 solar distances would be dramatic, and would reveal Earth's motion in a very short period of time. Nor does the Trapezium show any differential parallax, nor do the other close star groupings that Galileo observed. In fact, no double star that an astronomer of that time might have observed shows differential parallax. The witness had not been given. The absence of differential parallax requires the stars to be far more distant than the values Salviati promotes, even far more distant than the twelve-times-2160-solar-distances value that he suggested based on the period of Mars. Thus, the stars must also be far larger than he says, if the universe is heliocentric.[49] The star size question remains.

---

[46] Leos Ondra, "A New View of Mizar," *Sky & Telescope* 108 (2004), 72–75. Prior to Ondra's work, the first observation of a double star had been attributed to Riccioli. Ondra made his discovery by going through Galileo's observing notes. An extended version of the *Sky & Telescope* article is available on Ondra's web page: http://www.leosondra.cz/en/mizar/.

[47] All this is found in Galileo's notes, in Antonio Favaro, ed., *Le Opere di Galileo Galilei: Edizione Nazionale Sotto gli Auspicii di Sua Maesta` il re d'Italia* (Firenze, 1890-1909), vol. 3, pt. 2, 877. Galileo recorded the apparent diameters as being 6 seconds and 4 seconds—somewhat larger than the apparent diameters he gave for fixed stars in the *Dialogue* discussion covered in this paper. He recorded their separation as being 15 arc seconds, in complete agreement with modern measurements. Galileo's telescopic observations were remarkably accurate—see Christopher M. Graney, "On the accuracy of Galileo's observations," *Baltic Astronomy,* 15 (2007), 443-49.

[48] Harald Seibert, "The Early Search for Stellar Parallax: Galileo, Castelli, and Ramponi," *Journal for the History of Astronomy,* 36 (2005), 251–71.

Salviati's second suggestion for a method to detect Earth's motion involves the fixed stars together with a pole or wooden beam, a telescope, and a lot of distance. Imagine, he says, an open plain. On the north side of the plain is a mountain, atop which is a chapel, with a horizontal beam of wood or some other material mounted above its roof. Then, he says,

> I shall seek in the plain that place from which one of the stars of the Big Dipper is hidden by this beam… just when the star crosses the meridian. Or else, if the beam is not large enough to hide the star, I shall find the place from which the disc of the star is seen to be cut in half by the beam—an effect which can be discerned perfectly by means of a fine telescope.[50]

Thus as Earth moves, the star will (when observed under the same conditions at different dates) peek out on one side or the other of the obstructing beam—that is, exhibit a parallax—owing to the change of Earth's position relative to the star. Here Salviati is invoking the ability of the telescope to show the disk of a star "bare and very many times enlarged" that he mentioned earlier in the discussion. Figure 7 shows a representation of the appearance of a star as seen through a very small telescope such as Galileo's, with simulations of the beam cutting the star in half and revealing Earth's motion in the manner Salviati envisions.

There is something fundamentally problematic in what Salviati says here. As we shall see shortly, the disk-like appearance of stars in a small telescope, seen in Figure 4, was a false product of the telescope, formed entirely within the telescope. It did not exist outside the telescope. And since it did not exist outside the telescope, it could not be cut in half by anything outside the telescope. Any astronomer who tried in any way to cut the disk of a star by means of an external obstructing object would quickly realize that Salviati was talking nonsense about this supposedly perfectly discernable effect.

---

[49] Christopher M. Graney and Henry Sipes, "Regarding the Potential Impact of Double Star Observations on Conceptions of the Universe of Stars in the Early 17th Century," *Baltic Astronomy,* 18 (2009), 93-108.
[50] *Dialogue*, 451.

Salviatti's "obstructing beam" discourse is roughly the end of the *Dialogue's* star size discussion. He declares that the detection of parallax in a star would be a great achievement in astronomy, "for by this means, besides ascertaining the annual motion, we shall be able to gain a knowledge of the size and distance of that same star." Salviati and Sagredo exchange a few words about whether anyone has ever tried the obstructing beam procedure, with Salviati stating that he thinks not, "for it is improbable that if anyone had tried this he would not have mentioned the result, whichever opinion it turned out to favor."[51] Sagredo states his complete satisfaction, and moves to change the topic to details about Earth's motion. This ends Salviati's response to the star size argument from Scheiner and Locher's *Disquisitions*.

QUESTIONS RAISED BY A FALSE UNIVERSE

Tacquet's assessment that the *Dialogue's* long discourse on the star size question failed to elude or even weaken the heliocentric star size problem seems robust. Salviati's own numbers for stellar distances and apparent sizes point to even the smallest visible stars being at least five times the sun's actual diameter. His proposed methods for eventually detecting Earth's motion are problematic and undermine even those numbers. He cannot refute the compact argument Simplicio presents from Scheiner's *Disquisitions*. Remember, of course, that Kepler would not have thought Salviati should refute it; Kepler embraced a heliocentric universe of giant stars and a single unique sun (Figure 8). But Salviati tried to promote a heliocentric universe of sun-like stars, like the universe of Bruno that Kepler opposed. In this Salviati failed. To a knowledgeable and experienced astronomer, Salviati must have appeared to be the simpleton on this subject.

Tacquet offers no specific criticism of the *Dialogue*; was he sufficiently knowledgeable and experienced to fully recognize the depth of Salviati's problems? After all, Vanpaemel notes that Tacquet either made no telescopic observations himself or, if he did, no record of them survives.[52] However, Tacquet cites Riccioli. Riccioli certainly possessed that knowledge and experience. In his *New Almagest*, Riccioli writes of Mizar as being a double star, treats the subject of telescopically measured apparent stellar diameters at length, and calculates the actual sizes of stars required under a heliocentric universe given specific parallax limits (finding that even little Alcor must rival Earth's orbit in size)—and these things take up but a fraction of that book's 1400 pages of dense text and diagrams.  And *Disquisitions* and the *New Almagest* were not the only works by Jesuit astronomers.  Whether Tacquet himself had the knowledge and experience to fully recognize the shortcomings of Salviati's discourse, that knowledge and experience was available among Jesuit astronomers.

---

[51] *Dialogue*, 451.
[52] Vanpaemel, "Jesuit Science," 407.

The Scheiner-Galileo-Tacquet exchange illustrates that the heliocentric universe that Jesuit astronomers saw was Kepler's heliocentric universe, with its tiny, brilliant, unique sun orbited by a retinue of tinier planets and surrounded by a universe of distant, enormous, dim, and not-very-sun-like stars.  Works by Jesuit astronomers undoubtedly informed others in the Church, including those involved in the actions taken against heliocentrism in the first half of the seventeenth century.

A thorough investigation of the extent to which Kepler's universe and the star size question informed those actions remains to be undertaken.  Certainly not all opinions regarding heliocentrism were informed by the implications of carefully measured apparent star sizes and the failure to detect parallax.  Not everyone who favored the Copernican hypothesis considered the stars to be giant.  Not everyone who opposed that hypothesis did so because of the star size question and concerns about God and intellectual economy.  Bruno, of course, is an example of the former.  Niccolò Lorini, who in 1615 filed a complaint about Galileo with the Inquisition in Rome, would seem to be an example of the latter, as his complaint mentions not heliocentrism and star sizes but rather conflicts between heliocentrism and a plain reading of scripture.[53]

However, some persons in the Church who were involved in the actions taken against heliocentrism certainly were informed about the star size question.  Msgr. Francesco Ingoli, who Galileo believed to have been influential in the rejection of heliocentrism by the Congregation of the Index in 1616, cited the star size argument against Copernicus in his writings.[54]  Fr. Melchior Inchofer, S.J. who was selected for a three-person Special Commission formed by Pope Urban VIII to investigate the publication of the *Dialogue*, likewise noted the star size question.[55]

---

[53] Finocchiaro, *Galileo Affair*, 28, 134-35.
[54] Ingoli in a 1616 essay to Galileo referenced Brahe and noted that the Copernican system requires "the fixed stars to be of such size, as they may surpass or equal the size of the orbit circle of the Earth

Suppose familiarity with the star size question, illustrated in the Scheiner-Galileo-Tacquet exchange, went beyond Ingoli and Inchofer. Might we then look differently at a more prominent figure like Robert Cardinal Bellarmine and his interactions with heliocentrism and Galileo? Bellarmine famously demanded that hard evidence in favor of heliocentrism be produced prior to any re-interpretation of the Bible's references to Earth's immobility or the sun's motion as referring to only appearances rather than to actualities. Writing in April 1615, he said,

> I say that if there were a true demonstration… that the sun does not circle the earth but the earth circles the sun, then one would have to proceed with great care in explaining the Scriptures that appear contrary, and say rather that we do not understand them than that what is demonstrated is false. But I will not believe that there is such a demonstration, until it is shown me. Nor is it the same to demonstrate that by supposing the sun to be at the center and the earth in heaven one can save the appearances, and to demonstrate that in truth the sun is at the center and the earth in heaven; for I believe the first demonstration may be available, but I have very great doubts about the second, and in the case of doubt one must not abandon the Holy Scripture as interpreted by the Holy Fathers.[56]

Some have argued that Bellarmine was not truly interested in such evidence and would not have been persuaded by it. Consider, for example, the modern Jesuit astronomer George Coyne, Director of the Specola Vaticana, the Vatican's astronomical observatory at Castel Gandolfo in Italy and Mt. Graham in Arizona, from 1978 to 2006, and a member of the 1980s Galileo Commission appointed by Pope St. John Paul II. Citing Bellarmine's statement that "I will not believe that there is such a demonstration until it is shown me," Coyne wrote:

---

itself"—see Graney, *Setting Aside*, 70-72, 167-68; this book includes a full translation of Ingoli's essay.

[55] Inchofer in his 1633 *Tractatus Syllepticus* noted that "the defenders of the Copernican system imagine that, since the stars are seen at an almost infinite distance, they have a size which is explicable by hardly any proportion"—see Richard J. Blackwell, *Behind the Scenes at Galileo's Trial: Including the First English Translation of Melchior Inchofer's Tractatus Syllepticus* (Notre Dame, Indiana, 2006), 182.

[56] Finocchiaro, *Galileo Affair*, 68.

it is clear that Bellarmine was convinced that there was no such demonstration to be shown. A further indication of this conviction of Bellarmine is had in the fact that he supported the decree of the Congregation of the Index in 1616 [declaring heliocentrism both "false" and "contrary to the Holy Scripture"[57]] which was aimed at excluding any reconciliation of Copernicanism with Scripture. If Bellarmine truly believed that there might be a demonstration of Copernicanism, why did he not recommend waiting and not taking a stand?... And why did he accept to deliver an admonition to Galileo in 1616? This admonition prohibited Galileo from pursuing his research as regards Copernicanism. Galileo was forbidden to seek precisely those scientific demonstrations which, according to Bellarmine, would have driven theologians back to reinterpret Scripture.... Bellarmine was convinced that there would never be a demonstration of Copernicanism and that the Scriptures taught an Earth-centered universe.[58]

Consider also Harvard astronomer and historian of science Owen Gingerich, who has suggested that Bellarmine might not have accepted as a true demonstration of Earth's motion even the discovery of annual parallax, or of the Foucault Pendulum often displayed in science museums today as a demonstration of Earth's rotation.[59]

Other authors assess Bellarmine differently, of course. Historian David Wootton, for example, has written that Bellarmine was not advocating a ban on the discussion of heliocentrism and was prepared accept that it could be demonstrated to be true.[60] And Bellarmine himself had stated that evidence could indeed supersede Scripture in astronomical matters. Decades before he became involved with the debate over heliocentrism, he had argued in his Louvain lectures that Scripture suggested that celestial bodies were not carried by any celestial machinery, but rather moved autonomously, "like the birds of the air or the fish of the water." This was contrary to then-standard astronomy; Christopher Clavius, S.J., in his widely used astronomy text,

---

[57] Finocchiaro, *Galileo Affair*, 149.
[58] George V. Coyne, "Galileo and Bellarmine," in: *The Inspiration of Astronomical Phenomena VI: ASP Conference Series,* Vol. 441, ed. Enrico Maria Corsini (San Francisco, 2011), 7-8.
[59] Owen Gingerich, *God's Universe* (Cambridge, 2006), 91-94.
[60] David Wootton, *Galileo: Watcher of the Skies* (New Haven, 2013), 144-45.

Page 24 of 39

specifically said that celestial bodies do not move like birds or fish. However, Bellarmine granted that, while Scripture seemed to support autonomous motion,

> If then one ascertained with evidence that the motions of the heavenly bodies are not autonomous... one would have to consider a way of interpreting the Scriptures which would put them in agreement with the ascertained truth: for it is certain that the true meaning of Scripture cannot be in contrast with any other truth.[61]

Nevertheless, let us imagine the Bellarmine of 1615 as Coyne did—adamant that no evidence could demonstrate the veracity of heliocentrism. The heliocentric universe that Coyne's Bellarmine probably had in mind when he expressed skepticism toward heliocentrism in 1615, and eventually stood against it in 1616, would have been Kepler's giant stars universe. After all, that is what Bellarmine would have gathered from Jesuit astronomers like Scheiner. And if Salviati could not effectively argue against Scheiner in 1632, Galileo certainly could not have done so in 1615. Do we then look differently at even Coyne's Bellarmine? As he stood convinced that no demonstration of heliocentrism was possible, the heliocentrism he would surely have had in mind was a heliocentrism that even Sagredo would call "entirely unbelievable"—all while Brahe's geocentric universe with its stars located just beyond Saturn and commensurate in size with the other celestial bodies was a scientifically viable option, supported by Jesuit astronomers.

While the extent to which these sorts of astronomical considerations informed Bellarmine's view on heliocentrism awaits investigation, examples can be found of such considerations being thought capable of swaying the interpretation of Scripture. Bellarmine's Louvain lectures are one example. Another is an assessment in 1661 from the French Jesuit Honoré Fabri regarding heliocentrism specifically—that the Church does not operate against evidence, and so if some demonstration of the validity of the Copernican hypothesis were found, the Church would not scruple to declare that those passages of Scripture that speak of an immobile Earth or a moving sun are to be understood in a figurative sense.[62] A third is the Inquisition consultant and Jesuit Pietro Lazzari, who urged in 1757 that the general prohibition against "all books teaching the earth's motion and the sun's immobility" be removed from the *Index*. Lazzari said that the evidence was indeed against Copernicus in 1616 when the Congregation of the Index had declared heliocentrism "false" and contrary to Scripture:

> Thus, one can say that at that time there were good reasons or motives for [the Copernican system] being prudently prescribed. I consider three of these reasons for prescribing it. Firstly, this opinion of the earth's motion was new and was rejected and branded with serious objections by *most excellent astronomers and physicists*. Secondly, it was deemed to be contrary to Scripture when taken in the proper and literal sense; and this was conceded even by the defenders of that opinion. Thirdly, no strong reason or demonstration was advanced to oblige or counsel us to so disregard Scripture and support this opinion.[63]

---

[61] Roberto Bellarmino, *The Louvain lectures (Lectiones Lovanienses) of Bellarmine and the Autograph copy of his 1616 declaration to Galileo,* eds. Ugo Baldini and George V. Coyne (Città del Vaticano, 1984), 19-20, 38 n. 88.

[62] Maurice Finocchiaro, *Retrying Galileo, 1633-1992* (Berkeley, 2005), 93-94.

Lazzari goes on to note all the strong evidence that had since accumulated for heliocentrism, specifically since the advent of Newtonian physics at the end of the seventeenth century, that now obliged and counseled support for the heliocentric opinion.[64] Lazzari prevailed. The general prohibition against heliocentrism was lifted.

Beyond any specific figure such as Bellarmine, does the star size question illustrated in the Scheiner-Galileo-Tacquet exchange prompt us to look differently even at institutional actions like the prohibition against heliocentrism? Does it matter what sort of heliocentric universe the Congregation of the Index had in mind when it labelled heliocentrism "false"? What does it mean if the heliocentric universe they had in mind was Kepler's—the heliocentric universe that observations, measurements, and calculations required? After all, that universe was indeed "false," insofar as today astronomers know that the sun is not the unique central body of all the universe, and the stars are not all far larger than the sun. Modern astronomers see great diversity in the stars: a very small portion of the stars are indeed far larger than the sun; many more are comparable to the sun; the vast majority are small "red dwarf" stars far outclassed by the sun.

This paper suggests no answers to the questions proposed here. That is not its purpose. The purpose of this paper is to illustrate how, to Jesuit astronomers in the seventeenth century—and indeed to any astronomers of that time who were persuaded by observations, measurements, and calculations, and to anyone who might have consulted such astronomers for their expertise—the Copernican universe looked radically different from the universe as modern astronomers understand it. But understanding this allows these sorts of interesting questions to be asked. Thus the details of the Scheiner-Galileo-Tacquet exchange are worth the time of the general reader of *The Catholic Historical Review*.

---

[63] Finocchiaro, *Retrying Galileo*, 139 (italics added).
[64] Finocchiaro, *Retrying Galileo*, 137-48.

FALSE STARS

No other Jesuit could follow Tacquet in effectively wielding the star size argument against heliocentrism, however. In 1674, Robert Hooke published his *An Attempt to Prove the Motion of the Earth*. In it, he cites Tacquet as being a leading anti-Copernican who wielded the star size argument forcefully. But Hooke also mentions an observation that would in time nullify the star size argument. This was not the supposed observation of annual parallax that was the central feature of the book. No, Hooke reported that he had been able to observe through his telescope during the daylight a fixed star at the zenith; he noted how that star appeared to be very, very small; and he noted that this observation answered Tacquet (and Riccioli):

> [B]y this Observation of the Star in the day time when the Sun shined, with my 36 foot Glass I found the body of the Star so very small, that it was but some few thirds [i.e. *sixtieths* of an arc second] in Diameter, all the spurious rayes that do beard it in the night being cleerly shaved away, and the naked body thereof left a very small white point.
>
> The smalness of this body thus discovered does very fully answer a grand objection alledged by divers of the great *Anti-copernicans* with great vehemency and insulting; amonst which we may reckon *Ricciolus* and *Tacquet*, who would fain make the apparent Diameters of the Stars so big, as that the body of the Star should contain the great Orb [Earth's orbit] many times, which would indeed swell the Stars to a magnitude vastly bigger then the Sun, thereby hoping to make it seem so improbable, as to be rejected by all parties. But they that shall by this means examine the Diameter of the fixt Stars, will find them so very small that according to these distances and Parallax they will not much differ in magnitude from the body of the Sun....[65]

In fact, even Hooke's method did not reveal the naked body of a star—that remains beyond most telescopes even today—but his observation gave evidence that previous telescopic measurements of star sizes were in error. Nor was Hooke the first to obtain such evidence.

In fact, the problem of measuring the apparent diameter of a "star," be it a fixed star or a wandering "star"—that is, a planet—was most difficult. For example, consider the case of the wandering star Venus. A keen eye saw Venus as a bright dot with an apparent diameter approximately one tenth that of the moon (as Ptolemy and Brahe said). But through the telescope, Venus's disk appeared much smaller relative to the moon's disk—the telescope seemingly enlarged the moon more than it enlarged Venus (Figure 9). Moreover, seen through the telescope, the disk of Venus varied in apparent diameter over time, and showed the phases Galileo had discovered, ranging from nearly a full disk to a slim crescent (Figure 10). The explanation for all this was that the telescope stripped away the glare or "spurious rays" or "adventitious irradiation" from Venus, revealing the bare body of the wandering star, and thus its true form and its correct apparent size.

The telescope was thought to also do the same thing for fixed stars. The telescopes of the seventeenth century indeed revealed fixed stars to be distinct disks. Refer again to Figure 4, and consider the following observation recorded by John Flamsteed, the first English Astronomer Royal:

> 1672, October 22. When Mercury was about 10 deg. high, I observed him in the garden with my longer tube (of 14 foot); but could not with it see the fixa [fixed star] (near him), the daylight being too strong; only I noted his diameter 45 parts = 16″ [seconds of arc], or a little less; for, turning the tube to Sirius, I found his diameter 42 parts = 15″, which I judged equal to Mercury's. The aperture on the object-glass was ¾ of an inch: so that Sirius was well deprived of spurious rays,

---

[65] Robert Hooke, *An Attempt to Prove the Motion of the Earth from Observations* (London: Printed by T.R. for John Martyn, 1674), 26.

and shined not turbulently, but as sedate as Mercury; the limbs of both well
defined, but Sirius best.[66]

Note Flamsteed's indication that the disk of Sirius was more clearly defined than that of the planet Mercury. Note also that the apparent diameter of 15 seconds that Flamsteed gives for Sirius is not much different from Riccioli and Tacquet's value of 18 seconds. The telescope also improved sensitivity to parallax, forcing the stars to be more distant. Riccioli's calculation of star sizes that showed that even little Alcor must rival Earth's orbit in size was based on this improved sensitivity.[67]

Neither Flamsteed nor Riccioli nor any other astronomer of the time understood that, in the case of a fixed star, the disk revealed by the telescope was false and formed within the telescope itself, a product of "diffraction"—the interaction of light waves with the small aperture of the telescope. On the other hand, in the case of Venus, the disk (and its phases) revealed by the telescope was true. This was a most difficult issue, not to be fully worked out until a satisfactory wave theory of light was developed in the early nineteenth century.[68] At any rate, as Flamsteed's 1672 observation record shows, during much of the seventeenth century, the telescopically observed and measured disks of stars were thought to be the actual bodies of those stars.

Hooke obtained evidence that they were not, but he was not the first to do so. Shortly before Tacquet's death in 1660, Christiaan Huygens had published observations similar to Hooke's, although Flamsteed rejected them (indeed, Flamsteed's observations of Sirius and Mercury quoted above were part of a discussion rejecting Huygens). Shortly thereafter, however, Johannes Hevelius published observations by Jeremiah Horrocks to the effect that the disks of stars could not be cut by an obstructing body (the moon, specifically), which suggested they were spurious. Riccioli himself may have lost faith in the star size argument by the mid-1660s.[69] Hooke's work simply dealt another

---

[66] Graney, *Setting Aside*, 152.
[67] Graney, *Setting Aside*, 135.
[68] Christopher M. Graney and Timothy P. Grayson, "On the telescopic disks of stars – a review and analysis of stellar observations from the early 17th through the middle 19th centuries," *Annals of Science* 68 (2011), 351-73.
[69] Graney, *Setting Aside*, 148-157. There is also an intriguing discussion in Kepler's 1618 *Epitome of Copernican Astronomy*. Earlier in *Epitome* Kepler puts forward the same ideas as discussed here: stars being large (Figure 8); the idea that the farther away a star is supposed to be, the larger its true physical size must be; the use of star sizes against Bruno. But several hundred pages later, Kepler contradicts his earlier discussion by stating that stars are incredibly small—just a tiny

blow to the argument. The spurious nature of any telescopic measurement of apparent stellar diameters was becoming manifest. The idea that Salviati promoted, that stars were sun-like bodies at distances so vast that Earth's orbit was nothing by comparison, was becoming something that observations and measurements could actually support—that is, that *science* could actually support. But science did not progress steadily in this case. In 1717 Jacques Cassini, a well-respected French astronomer, published a paper supporting the idea that telescopic measurement of apparent stellar diameters were indeed valid. He republished this idea in a book of 1740.[70] When Lazzari was arguing for the Copernican system in 1757 one could find a recently-published astronomy text that, citing Cassini, reported how a telescope could reliably reveal the apparent diameter of bright fixed stars to be about five seconds of arc, just as Salviati claimed more than a century earlier.[71] Firm scientific support for the stars being sun-like bodies did not come easily.

---

    fraction of the diameter of the sun. This stems from an argument about the nature of the universe that is based not on observations but on assumptions Kepler makes regarding matter and the densities of different regions of the universe. Kepler justifies these tiny stars by means of a general statement that the fixed stars, when observed by a skilled astronomer using a telescope, appear as mere points of light. This raises the possibility that Kepler had learned of observations suggesting the false nature of telescopically measured star sizes well before Horrocks and Huygens. Likewise, Ingoli makes an intriguing comment about stars not "operating" like other bodies. However, one telescopic astronomer after another in the first two thirds of the seventeenth century referred to stars seen through telescopes as being finite, measurable disks, and not as being immeasurable points. Thus, absent a description such as that provided by Horrocks or Huygens that explains how a skilled astronomer would see stars as mere points, or how stars would operate differently, Kepler's "points" comment and Ingoli's "operating" comment should perhaps be considered to be anomalous. See Graney, "Starry Universe," 162-66.

[70] Jacques Cassini, "De le Grandeur des Etoiles Fixes, et de Leur Distance a la Terre," 13 Nov. 1717, in: *Histoire de l'Académie Royale des Sciences, Avec les Mémoires de Mathématique et de Physique Annee MDCCXVII* (Paris, 1741). Cassini describes measuring the apparent diameter of Sirius by observing both it and Jupiter with a long telescope of reduced aperture, and comparing the two bodies to find that Sirius measures 1/10$^{th}$ Jupiter's diameter, or about five seconds: "Le diametre apparent de Jupiter étoit alors de 50 secondes, d'où il resulte que celui de Sirius étoit d'environ 5 secondes [pp. 258-59]." The same can be found in found in Cassini's book *Elements d'Astronomie* (Paris, 1740), 50.

[71] John Hill, *Urania, or a Compleat View of the Heavens, etc. in the form of a Dictionary,* Vol. I (London, 1754): In the entry for "STARS, fixed" (no page numbers) is "The observation of Sirius's diameter being five seconds, had, for its author, one of the most accurate, and most judicious astronomers the world has ever known, Cassini, and, *whenever it is repeated with the same apparatus, it succeeds in the same manner, and verified very punctually* [italics added]." The author also discusses the alternate view that stars do not have truly measurable apparent diameters, stating "it may be well to advise repeated experiments and observations farther to determine which is right." The entry for "DISTANCE of the fixed Stars" describes Cassini's method of measuring stellar apparent diameters, and it also gives five seconds as the diameter of Sirius, and states that those who grant the most prominent fixed stars no apparent diameters seem to be "carrying it too far."

CONCLUSION: SCIENCE, STARS AND UNIVERSES

Modern astronomers indeed see great diversity in the stars, but in a broad sense modern astronomers do understand stars to be sun-like bodies at vast distances. Thus it might seem to follow that those voices from four centuries ago that promoted the idea that stars were distant suns—whether they were the voices of real persons like Bruno, or of fictional characters like Salviati—were "voices of science" in some sense. Correspondingly, it might seem that opposing voices from that time were in some sense in opposition to science, disinterested in scientific evidence.[72]

But through much of the seventeenth century, science—observations measurements, calculations—did not support that modern view of what stars are. The views science supported were Kepler's heliocentric view of stars being distant, dim bodies that utterly dwarf the sun and the other celestial bodies, and Brahe's geocentric view of stars being just beyond Saturn and commensurate in size with the other celestial bodies. Jesuit astronomers like Christoph Scheiner, Giovanni Battista Riccioli, and André Tacquet understood this. Others who might have listened to Jesuit astronomers, including those involved in actions by Church authorities against Galileo and the heliocentric system, would have understood this, too.

To many, Kepler's starry universe was what Galileo's character Sagredo said—"beyond belief," a violation of intellectual economy, or even perhaps, "foolish and absurd in philosophy." Brahe's geocentric universe was the remaining choice. Tacquet says that Galileo tried in vain to elude, in essence, the choice of either Kepler or Brahe. Tacquet sees that while he himself and other anti-Copernicans could offer science against a heliocentric universe of sun-like stars, Galileo's character Salviati could offer only intuition in favor of such a universe.

That intuition turned out, of course, to be correct. But at the time that intuition could not survive even the comments of Galileo's character Simplicio, much less Jesuit astronomers like Scheiner and Tacquet wielding observations and calculations against it. Thus the back-and-forth from Scheiner to Galileo to Tacquet shows that the heliocentric universe that observations and measurements required, that astronomers discussed and

---

[72] See Mario Livio, *Galileo and the Science Deniers* (New York, 2020).

Church authorities condemned—and that Simplicio rejected, that Sagredo said was beyond belief, and that Salviati sought in vain to elude—was Kepler's universe of monstrous stars enveloping one single, unique solar system and its tiny, brilliant sun. This universe looked radically different from the universe as we understand it today. No modern view of a universe of many distant suns would be scientifically supportable until after Tacquet's death in 1660.

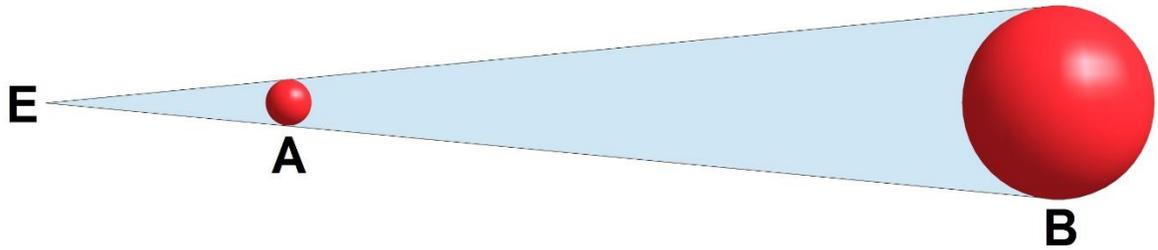

Figure 1. The more distant an observed object is from Earth (E), the greater its actual physical size must be to retain the same apparent size (indicated by the shaded region). A and B will each have the same apparent size as seen from E, but B is over four times more distant from E than is A, so B has over 4 times the actual diameter of A, and over 64 times (4 cubed) the volume of A.

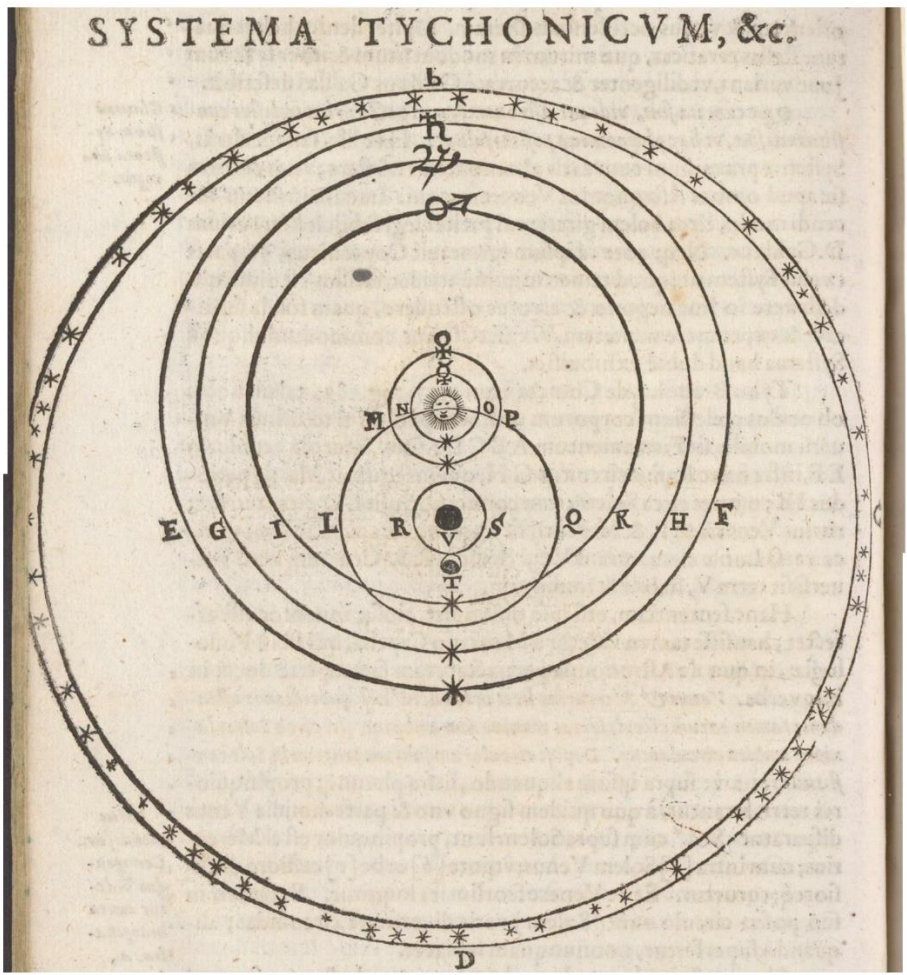

Figure 2. Tycho Brahe proposed that the sun, moon, and fixed stars circle the Earth while the planets circle the sun.  Brahe's system and Copernicus's system were identical insofar as any Earth-bound observations of sun, moon, and planets were concerned.  As Johannes Kepler said, "in Brahe the Earth occupies at any time the same place that Copernicus gives it, if not in the very vast and measureless region of the fixed stars, at least in the system of the planetary world."[73]  Thus Brahe's system was fully compatible with telescopic discoveries involving those bodies.  This illustration of Brahe's system is from the *Mathematical Disquisitions* of Scheiner and Locher.  Image credit: ETH-Bibliothek Zürich, Alte und Seltene Drucke.

---

[73] Johannes Kepler, *Epitome of Copernican Astronomy and Harmonies of the World*, trans. Charles Glenn Wallis [Great Minds Series]  (Amherst, NY, 1995), 175.

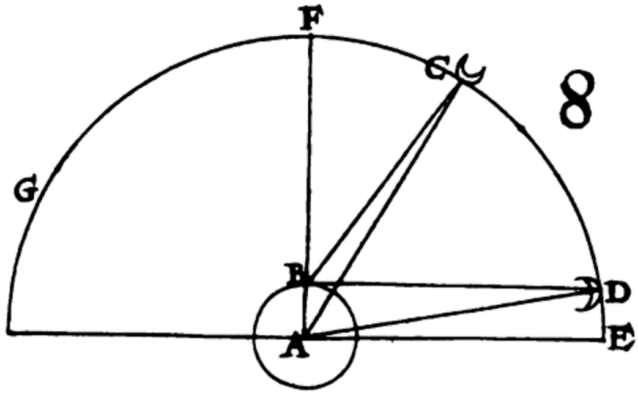

Figure 3. Diagram in *Opera Omnia* for Tacquet, Book 5, Chapter 2, Number 21, showing the use of triangulation from Earth's surface to determine the distance to a celestial body (the moon, it would seem). Lines BC and BD show the position, as seen from B, of the moon when it is high in the sky and low on the horizon, respectively. Lines AC and AD are its position seen from points on Earth directly below it, or from the center of the Earth. With angles measured, trigonometry reveals the distance to the moon. Were the Earth much smaller (or the moon much more distant) the angles BCA and BDA would be too small to determine. Image credit: Google Books.

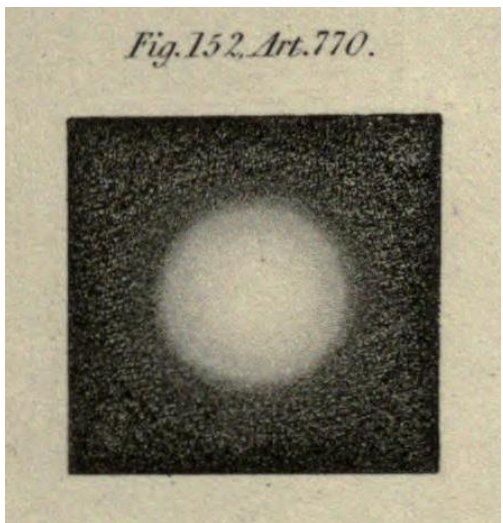

Figure 4. Illustration from John F. W. Herschel's 1828 *Treatises on Physical Astronomy, Light and Sound Contributed to the Encyclopædia Metropolitana* (p. 491 and plate 9) of a star as seen through a telescope of aperture similar to what was used for stellar observations in the seventeenth century. Seventeenth-century astronomers thought that a telescope stripped stars of "spurious" or "adventitious" rays, so that what was seen here was, as Galileo put it, the bare star. In fact, the "diffraction" of light waves through the telescope's aperture creates the globe-like appearance seen here, greatly inflating the star's apparent size. What is seen here is in fact spurious, and not the bare body of the star. A full understanding of diffraction, the wave nature of light, and star images seen in telescopes was not developed until the early nineteenth century. Image credit: ETH-Bibliothek Zürich, Alte und Seltene Drucke.

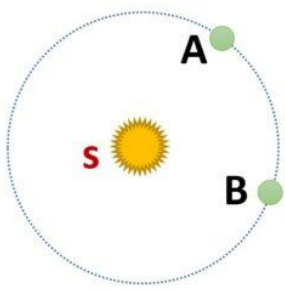

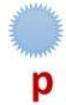

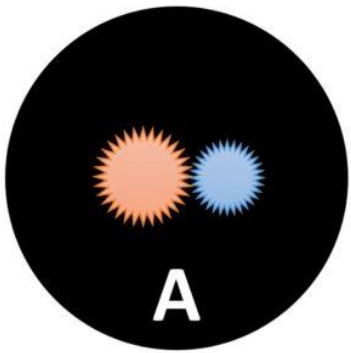
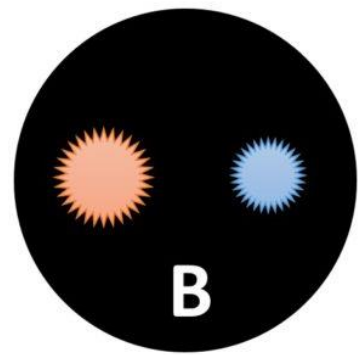

Figure 5. Earth orbits the sun (s), as shown at top left, moving clockwise from A to B. Stars q and p are both similar in size to the sun, located at differing distances from Earth. When the Earth is at A, the line of sight from Earth to q and p causes the two stars to appear very close together (below left). When Earth is at B, the two stars appear more separated (below right). When Earth returns to A, the two stars will again appear very close together. This is "differential parallax."

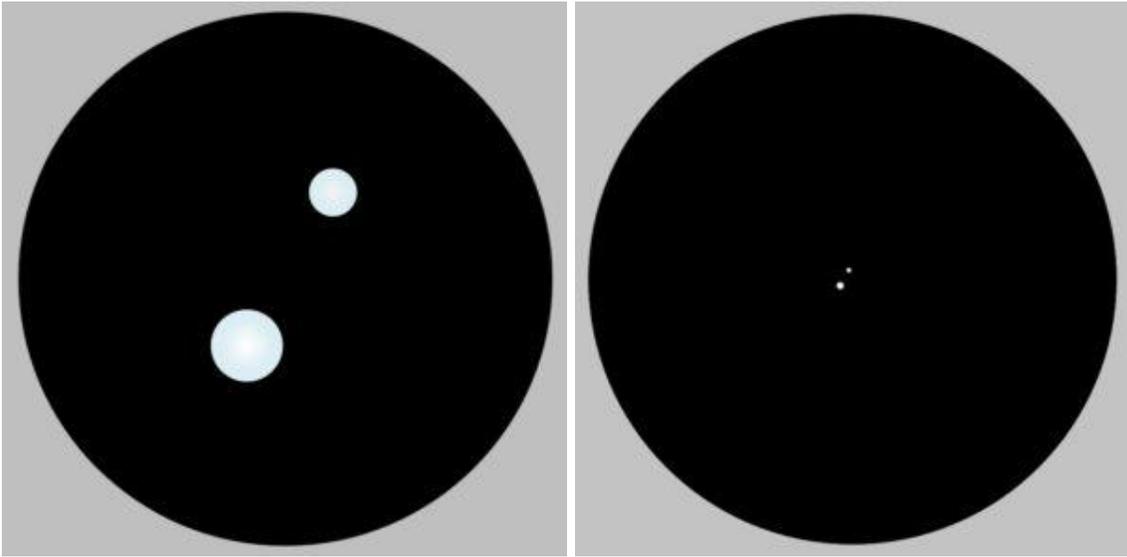

Figure 6. Left is the appearance of Mizar through Galileo's telescope, according to Galileo's measurements. The illustration at right perhaps gives a better idea of the view through Galileo's telescope.

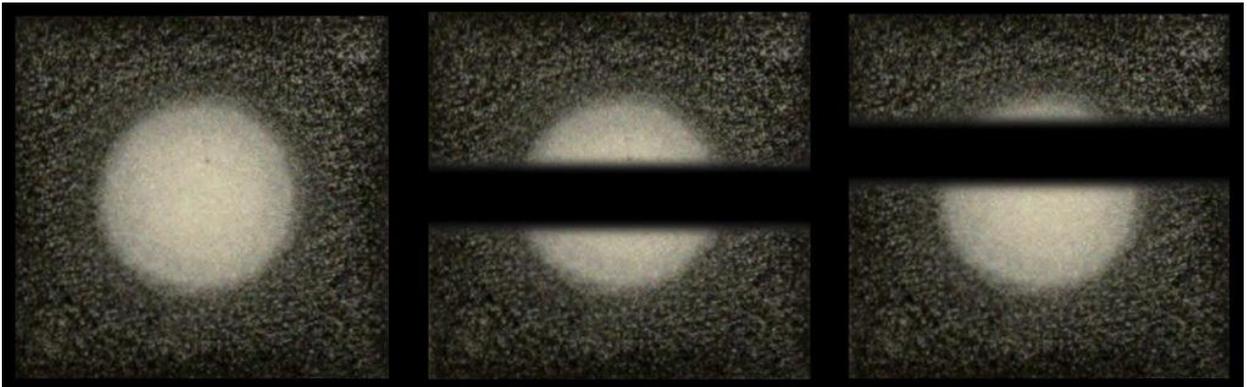

Figure 7. Left—a star seen through a telescope of very small aperture, from Figure 4. Center—simulated view of the star supposedly "cut in half" by Salviati's distant wooden beam. Right—simulated view showing how, after a period of time, the Earth's motion relative to the star might cause the position of the beam against the star to change, proving that Earth indeed moves. If, after one year, the star is once again divided in half by the beam, then Earth's motion around the sun (in which it returns annually to the same place) will be clearly demonstrated. As seen in Figure 4, the appearance of a star in a small telescope is entirely spurious. Thus this disk of the star, being but an artefact of light formed within the telescope, in fact cannot be cut by an external object in this manner.

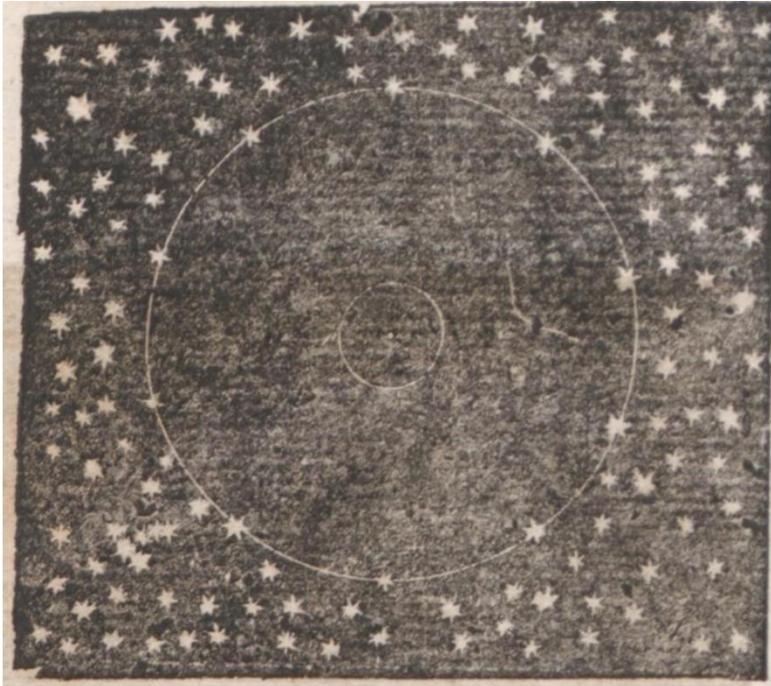

Figure 8. Diagram from Kepler's 1618 *Epitome of Copernican Astronomy* (p. 36), showing the sun as a small dot (just visible at the center), surrounded by larger stars. Image credit: ETH-Bibliothek Zürich, Alte und Seltene Drucke.

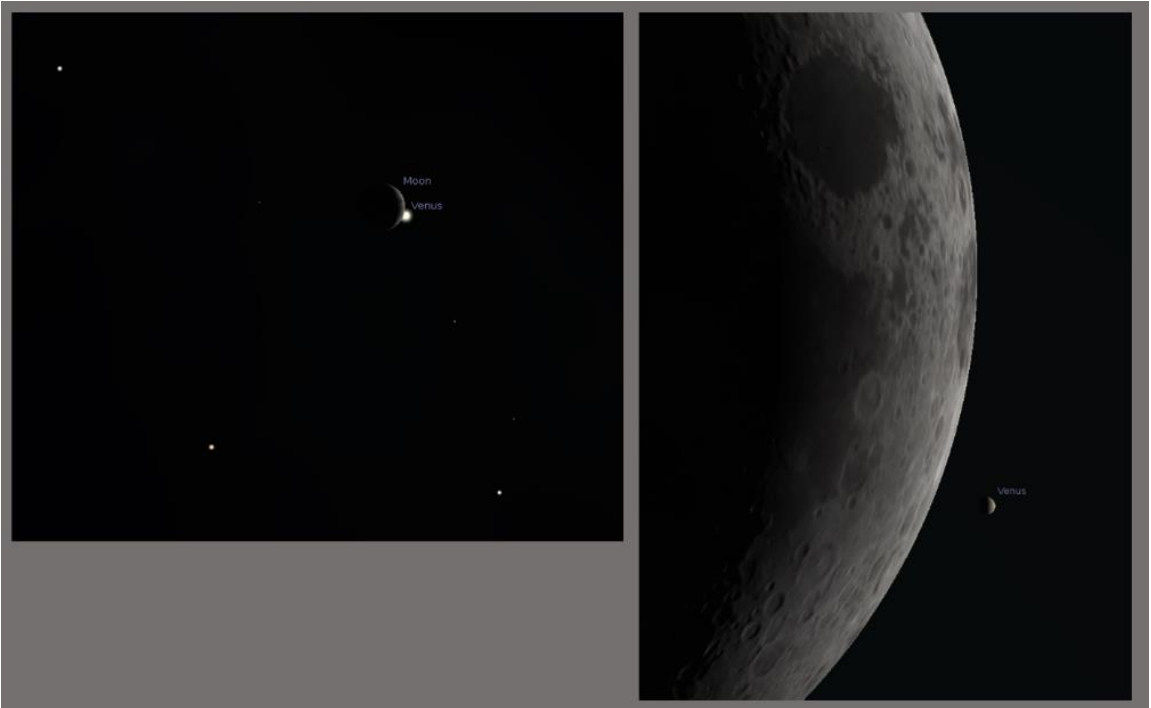

Figure 9. The moon and Venus, seen from the Kamchatka peninsula in November of 2021, as simulated via the *Stellarium* planetarium app. Left is the view with the eye; right is the view with the telescope. Note that the size of Venus relative to the moon is much smaller in the right-hand image, and that Venus shows a phase like the moon.

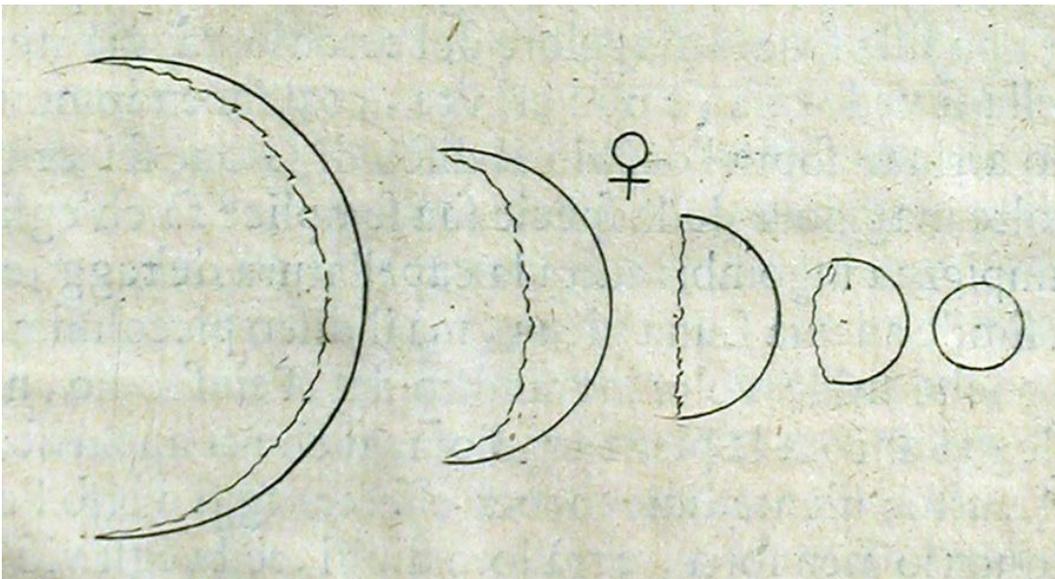

Figure 10. Galileo's illustration of the changing phases and apparent diameter of Venus, from his 1623 *Il Saggiatore* (p. 217). Image credit: History of Science Collections, University of Oklahoma Libraries.